\documentclass[12pt]{article}
\pdfoutput=1
\usepackage{amsmath}
\def\be{\begin{equation}}
\def\ee{\end{equation}}
\def\arr{\begin{array}{rll}}
\def\ea{\end{array}}
\def\bea{\begin{eqnarray}}
\def\eea{\end{eqnarray}}

\def\N2{$N{=}2$}

\def\>{\rangle}
\def\<{\langle}
\def\+{\dagger}
\def\={\ =\ }

\begin{document}
\renewcommand{\thefootnote}{\fnsymbol{footnote}}
\begin{titlepage}
\setcounter{page}{0}
\vskip 1cm
\begin{center}
{\LARGE\bf  $\mathcal{N}=2$ supersymmetric extensions  }\\
\vskip 0.5cm
{\LARGE\bf of relativistic Toda lattice}\\
\vskip 1cm
$
\textrm{\Large Anton Galajinsky \ }
$
\vskip 0.7cm
{\it
Tomsk Polytechnic University,
634050 Tomsk, Lenin Ave. 30, Russia} \\
{e-mail: galajin@tpu.ru}

\end{center}
\vskip 1cm
\begin{abstract} \noindent
$\mathcal{N}=2$ supersymmetric extensions of both the periodic and non--periodic relativistic Toda lattice are built within the framework of the Hamiltonian formalism. A geodesic description in terms of a non--metric connection is discussed.
\end{abstract}


\end{titlepage}

\renewcommand{\thefootnote}{\arabic{footnote}}
\setcounter{footnote}0

\noindent
{\bf 1. Introduction}\\

\noindent
The range of physical applications of the Calogero and Toda integrable systems is excessively broad. It encompasses the fractional statistics, quantum Hall effect, soliton theory, matrix models, supersymmetric gauge theories, and black hole physics.

It is known since the work of Ruijsenaars and Schneider \cite{RS,R} that both the Calogero and Toda models can be viewed as the non--relativistic limit of a more general integrable system, which enjoys the Poincar\'e symmetry realized in $1+1$ dimensions. In contrast to the non--relativistic theories, the Ruijsenaars--Schneider systems are described by the equations of motion which involve particle velocity.

While the non--relativistic Calogero and Toda models received tremendous attention in the past, their relativistic counterparts appear to be less popular. There are several reasons to focus on them more intently, though. Firstly, a geometric formulation underlying such systems is still missing. Whereas the non--relativistic models can be consistently embedded into the null geodesics of a Brinkmann--type metric \cite{G1,CG}, a similar description of the Ruijsenaars--Schneider systems seems problematic. For one thing, the Hamiltonian does not have a conventional quadratic form and the integrals of motion are not polynomial in momenta. For another, even if one is able to rewrite the equations of motion in the geodesic form \cite{ABHL}, one reveals a non--metric connection \cite{G}.\footnote{The rational variant of the Ruijsenaars-Schneider model can be linked to a metric connection. Yet, the geodesic motion actually takes place in a {\it flat} space parametrized by special curvilinear coordinates \cite{ABHL}.} Secondly, although the thermodynamic limit of the non--relativistic models is well understood (see, e.g., \cite{CLSTV,P} and references therein), an exhaustive analysis of the relativistic counterparts is still lacking. Thirdly, an important aspect of the studies over the past two decades has been the construction of supersymmetric extensions (for a review see \cite{FIL} and references therein). Yet, supersymmetric generalizations of the relativistic many--body models remain almost completely unexplored.

An $\mathcal{N}=2$ supersymmetric extension of the quantum trigonometric Ruijsenaars--Schneider model was built in \cite{BDM}. The corresponding eigenfunctions were linked to the Macdonald superpolynomials. A peculiar feature of the construction is that the fermionic operators and their adjoints obey the non--standard anticommutation relations, which reduce to the conventional ones in the non--relativistic limit only. The Hermitian conjugation of the fermions is realized in the non--standard fashion as well. In Ref. \cite{G}, the Hamiltonian methods were used to construct $\mathcal{N}=2$ supersymmetric extensions of the rational and hyperbolic three--body Ruijsenaars-Schneider models. A variant of the rational Ruijsenaars--Schneider model enjoying an arbitrary even number of supersymmetries and involving extra fermionic degrees of freedom was proposed in  \cite{KLS}. It is worth recalling, though, that the rational model describes a free system in disguise \cite{ABHL}.

The goal of this work is to extend our recent analysis in \cite{G} to the case of the relativistic Toda lattice \cite{R}. By making use of the Hamiltonian on--shell formalism, below we construct $\mathcal{N}=2$ supersymmetric extensions of both the periodic and non--periodic relativistic Toda lattice. In contrast to \cite{G}, the extension proves feasible for an arbitrary number of particles.

The work is organized as follows.

In Sect. 2, $\mathcal{N}=2$ supersymmetric generalizations of the periodic Toda lattice are built. We start with a positive--definite Hamiltonian and represent it as the sum of squares of structure functions, which obey a non--linear algebra. $\mathcal{N}=2$ supersymmetry charges are introduced in the conventional (cubic) polynomial form, the leading order of which is related to the structure functions specifying the bosonic Hamiltonian. Imposing the commutation relations of the $\mathcal{N}=2$ supersymmetry algebra, we obtain a set of partial differential equations to fix the bosonic functions entering the fermionic cubic terms in the supercharges. Two explicit solutions are found which generate consistent $\mathcal{N}=2$ supersymmetric extensions. It is known that the relativistic Toda lattice admits more than one Hamiltonian formulation (see, e.g., \cite{S}). We then consider an alternative Hamiltonian and build two more $\mathcal{N}=2$ supersymmetric generalizations.
In Sect. 3, the analysis is repeated for the non--periodic Toda lattice revealing four $\mathcal{N}=2$ extensions. In Sect. 4, the equations of motion of the relativistic Toda lattice are rewritten in the geodesic form. It is argued that the corresponding connection fails to be derivable from a metric. Some final remarks are gathered in the concluding Sect. 5.

Throughout the paper no summation over repeated indices is understood.

\vspace{0.5cm}

\noindent
{\bf 2. $\mathcal{N}=2$ supersymmetric extensions of periodic relativistic  Toda lattice}\\

\noindent
The relativistic Toda lattice is described by the equations of motion \cite{R}
\be\label{TL}
\ddot{x}_i={\dot x}_{i+1} {\dot x}_i W(x_{i+1}-x_i)-{\dot x}_i {\dot x}_{i-1} W(x_i-x_{i-1}), \qquad W(x-y)=\frac{g^2 e^{x-y}}{1+g^2 e^{x-y}},
\ee
where $i=1,\dots,N$ and $g$ is a coupling constant. The periodic case is characterized by the boundary conditions
\be\label{BC}
x_0=x_N, \qquad x_{N+1}=x_1.
\ee

As the first step in constructing an $\mathcal{N}=2$ supersymmetric extension, one introduces the momenta $p_i$ canonically conjugate to the configuration space variables $x_i$ and imposes the Poisson brackets
\be
\{x_i,p_j\}=\delta_{i,j},
\ee
where $\delta_{i,j}$ designates the Kronecker delta. The boundary conditions (\ref{BC}) imply the relations
\be\label{sr}
\{x_{i+1},p_j \}=\delta_{i+1,j}+\delta_{i,N}\delta_{j,1}, \qquad \{x_{i-1},p_j \}=\delta_{i-1,j}+\delta_{i,1}\delta_{j,N},
\ee
which are then used to verify that the positive definite Hamiltonian
\be
H_B=\sum_{i=1}^N e^{p_i}\left(1+g^2 e^{x_{i+1}-x_i}\right):=\sum_{i=1}^N \lambda_i \lambda_i, \qquad \lambda_i=e^{\frac{p_i}{2}}\sqrt{1+g^2 e^{x_{i+1}-x_i}}
\ee
does reproduce (\ref{TL}). The structure functions $\lambda_i$ prove to obey the non--linear algebra
\be\label{All}
\{\lambda_i,\lambda_j\}=\frac 14 \lambda_i \lambda_j \left( W(x_{i+1}-x_i)[\delta_{i+1,j}+\delta_{i,N}\delta_{j,1}]- W(x_{j+1}-x_j)[\delta_{i,j+1}+\delta_{i,1}\delta_{j,N}]\right).
\ee

As the second step, complex fermionic variables $\psi_i$, ${(\psi_i)}^{*}=\bar\psi_i$, $i=1,\dots,N$, are introduced which obey the canonical brackets
\be\label{FB}
\{\psi_i,\psi_j\}=0, \qquad \{\psi_i,\bar\psi_j\}=-{\rm i} \delta_{i,j}, \qquad \{\bar\psi_i,\bar\psi_j\}=0.
\ee
They allow one to build the Hamiltonian and the supersymmetry charges in the polynomial form \cite{G}
\bea\label{SUSY}
&&
Q=\sum_{i=1}^N \lambda_i \psi_i+{\rm i} \sum_{i,j,k=1}^N f_{ijk} \psi_i \psi_j {\bar\psi}_k, \qquad \qquad \qquad \quad \bar Q=\sum_{i=1}^N \lambda_i \bar\psi_i+{\rm i} \sum_{i,j,k=1}^N f_{ijk} \bar\psi_i \bar\psi_j \psi_k,
\nonumber\\[6pt]
&&
H=H_B-2{\rm i} \sum_{i,j,k=1}^N (f_{ijk}+f_{kji}+f_{ikj})\lambda_k \psi_i \bar\psi_j+{\rm i}\sum_{i,j,k,l,m,n=1}^N \{f_{ijl},f_{mnk} \}\psi_i \psi_j \psi_k \bar\psi_l \bar\psi_m \bar\psi_n
\nonumber\\[4pt]
&&
\qquad -\sum_{i,j,k,l=1}^N (\{\lambda_i,f_{klj}\}-\{\lambda_l,f_{ijk}\}+f_{ijp}f_{klp}-4f_{pil}f_{pkj})\psi_i \psi_j \bar\psi_k \bar\psi_l,
\eea
where $f_{ijk}=-f_{jik}$ are real functions to be fixed below. Finally, one verifies that the generators (\ref{SUSY}) obey the commutation relations of the $\mathcal{N}=2$ supersymmetry algebra
\be
\{Q,Q\}=0, \qquad \{Q,\bar Q \}=-{\rm i} H, \qquad \{\bar Q,\bar Q\}=0
\ee
provided the restrictions
\bea\label{EF}
&&
\{\lambda_i,\lambda_j \}+2 \sum_{k=1}^N f_{ijk} \lambda_k=0, \quad \{ \lambda_{\underline{k}} ,f_{\underline{n} \underline{m} l} \}+2 \sum_{p=1}^N f_{\underline{k} \underline{n} p} f_{p \underline{m} l}=0, \quad \{f_{{\underline{a}} \underline{b} \overline{c}}, f_{\underline{m} \underline{n} \overline{k}} \}=0
\nonumber\\[2pt]
&&
\eea
hold. In the previous formulae the underline/overline mark signifies antisymmetrization of the respective indices.

Comparing (\ref{All}) with the leftmost equation in (\ref{EF}), one gets
\bea\label{f}
&&
f_{ijk}=
\frac{1}{16} W(x_{j+1}-x_j)[\delta_{i,j+1}+\delta_{i,1}\delta_{j,N}][a \delta_{i,k} \lambda_j+(2-a)\delta_{j,k}\lambda_i]
\nonumber\\[2pt]
&&
\quad \quad  -\frac{1}{16}  W(x_{i+1}-x_i)[\delta_{i+1,j}+\delta_{i,N}\delta_{j,1}][(2-a)\delta_{i,k} \lambda_j+a\delta_{j,k}\lambda_i],
\eea
where $a$ is an arbitrary real constant. The second equation in (\ref{EF}) yields
\be\label{sup}
a(2-a)=0 \quad \Rightarrow \quad a=0 \quad \mbox{or} \quad a=2,
\ee
while the third equation in (\ref{EF}) turns out to be satisfied identically. Note that, in contrast to the $\mathcal{N}=2$ supersymmetric Ruijsenaars-Schneider systems studied in \cite{G}, the two options in (\ref{sup}) do not seem to be linked to one another by relabeling the (super)particles.

Given the explicit form of $f_{ijk}$ in (\ref{f}), (\ref{sup}), one can finally verify that the six--fermion term entering the Hamiltonian (\ref{SUSY}) is equal to zero. The model thus exhibits the properties of the conventional $\mathcal{N}=2$ supersymmetric many--body mechanics in which the supersymmetry charges are at most cubic in the odd variables, while the Hamiltonian is at most quartic in the fermions.

It is known that the relativistic Toda lattice admits more than one Hamiltonian formulation (see, e.g., the discussion in \cite{S}). Let us consider an alternative which is given by
\be
{\tilde H}_B=\sum_{i=1}^N e^{-p_i}\left(1+g^2 e^{x_{i}-x_{i-1}}\right):=\sum_{i=1}^N \tilde\lambda_i \tilde\lambda_i, \qquad \tilde\lambda_i=e^{-\frac{p_i}{2}}\sqrt{1+g^2 e^{x_{i}-x_{i-1}}}.
\ee
The structure functions $\tilde\lambda_i$ satisfy the algebra
\be\label{All1}
\{\tilde\lambda_i,\tilde\lambda_j\}=\frac 14 \tilde\lambda_i \tilde\lambda_j \left( W(x_i-x_{i-1})[\delta_{i-1,j}+\delta_{i,1}\delta_{j,N}]- W(x_j-x_{j-1})[\delta_{i,j-1}+\delta_{i,N}\delta_{j,1}]\right),
\ee
and, similarly to the analysis above, give rise to the phase space functions
\bea\label{ff1}
&&
{\tilde f}_{ijk}=
\frac{1}{16} W(x_j-x_{j-1})[\delta_{i,j-1}+\delta_{i,N}\delta_{j,1}][a \delta_{i,k} \tilde\lambda_j+(2-a)\delta_{j,k}\tilde\lambda_i]
\nonumber\\[2pt]
&&
\quad \quad  -\frac{1}{16}  W(x_i-x_{i-1})[\delta_{i-1,j}+\delta_{i,1}\delta_{j,N}][(2-a)\delta_{i,k} \tilde\lambda_j+a\delta_{j,k}\tilde\lambda_i],
\eea
which solve the restrictions (\ref{EF}), provided
\be
a=0 \quad \mbox{or} \quad a=2.
\ee
These add two more $\mathcal{N}=2$ models to the list above. At the moment, it is not clear whether the formulations based upon $(\lambda_i,f_{ijk})$ and $(\tilde\lambda_i,{\tilde f}_{ijk})$ can be connected with one another by a coordinate transformation.

\vspace{0.5cm}

\noindent
{\bf 3. $\mathcal{N}=2$ supersymmetric extensions of non--periodic relativistic Toda lattice}\\

\noindent
The non--periodic relativistic Toda lattice is obtained by imposing the boundary conditions
\be
x_0=\infty, \qquad x_{N+1}=-\infty,
\ee
which bring the equations (\ref{TL}) to the form
\bea\label{EOM}
&&
{\ddot x}_1={\dot x}_2 {\dot x}_1 W(x_2-x_1),
\nonumber\\[2pt]
&&
{\ddot x}_N=-{\dot x}_N {\dot x}_{N-1} W(x_N-x_{N-1}),
\nonumber\\[2pt]
&&
\ddot{x}_k={\dot x}_{k+1} {\dot x}_k W(x_{k+1}-x_k)-{\dot x}_k {\dot x}_{k-1} W(x_k-x_{k-1}),
\eea
where $k=2,\dots,N-1$ and, as before, $W(x-y)=\frac{g^2 e^{x-y}}{1+g^2 e^{x-y}}$.

Like in the preceding case, the Hamiltonian reproducing (\ref{EOM}) is the sum of squares of the structure functions $\lambda_i$
\bea\label{npT}
&&
H_B=\sum_{i=1}^N \lambda_i \lambda_i,
\nonumber\\[2pt]
&&
\lambda_N=e^{\frac{p_N}{2}}\sqrt{1+g^2 e^{-x_N}}, \qquad
\lambda_k=e^{\frac{p_k}{2}}\sqrt{1+g^2 e^{x_{k+1}-x_k}}, \qquad k=1,\dots,N-1,
\eea
which obey the non--linear algebra
\bea
&&
\{\lambda_i,\lambda_j\}=\frac 14 \lambda_i \lambda_j \left[ W(x_{i+1}-x_i) \delta_{i+1,j}-W(x_{j+1}-x_j) \delta_{i,j+1}\right],
\eea
with $i,j=1,\dots,N$. Note that the boundary conditions imply the standard Poisson bracket $\{x_i,p_j\}=\delta_{i,j}$ and the relations similar to (\ref{sr}) do not occur for the case at hand.

The construction of $\mathcal{N}=2$ supersymmetric extensions proceeds as above. Given $\lambda_i$ in (\ref{npT}), it suffices to construct $f_{ijk}$ which solve the master equations (\ref{EF}). From the leftmost condition in (\ref{EF}) one finds
\bea\label{f1}
&&
f_{ijk}=
\frac{1}{16} W(x_{j+1}-x_j) \delta_{j+1,i}[a \delta_{i,k} \lambda_j+(2-a)\delta_{j,k}\lambda_i]
\nonumber\\[2pt]
&&
\quad \quad  -\frac{1}{16}  W(x_{i+1}-x_i) \delta_{i+1,j}[(2-a)\delta_{i,k} \lambda_j+a\delta_{j,k}\lambda_i],
\eea
where $a$ is an arbitrary real constant. The second equation in (\ref{EF}) reveals two options
\be
a=0 \quad \mbox{or} \quad a=2.
\ee
The last constraint in (\ref{EF}) turns out to be satisfied identically. Similarly to the periodic case, one can verify that the six--fermion term entering the Hamiltonian (\ref{SUSY})
is zero for $f_{ijk}$ displayed in (\ref{f1}).

Concluding this section, let us discuss $\mathcal{N}=2$ supersymmetric extensions associated with the alternative Hamiltonian formulation based upon
\bea\label{npT1}
&&
{\tilde H}_B=\sum_{i=1}^N \tilde\lambda_i \tilde\lambda_i,
\nonumber\\[2pt]
&&
\tilde\lambda_1=e^{-\frac{p_1}{2}}\sqrt{1+g^2 e^{x_1}}, \qquad
\tilde\lambda_k=e^{-\frac{p_k}{2}}\sqrt{1+g^2 e^{x_{k}-x_{k-1}}}, \qquad k=2,\dots,N.
\eea
In this case the structure functions satisfy
\be\label{All1}
\{\tilde\lambda_i,\tilde\lambda_j\}=\frac 14 \tilde\lambda_i \tilde\lambda_j \left( W(x_i-x_{i-1}) \delta_{i-1,j}- W(x_j-x_{j-1}) \delta_{i,j-1}\right).
\ee
They give rise to
\bea\label{ff1}
&&
{\tilde f}_{ijk}=
\frac{1}{16} W(x_j-x_{j-1}) \delta_{i,j-1} [a \delta_{i,k} \tilde\lambda_j+(2-a)\delta_{j,k}\tilde\lambda_i]
\nonumber\\[2pt]
&&
\quad \quad  -\frac{1}{16}  W(x_i-x_{i-1}) \delta_{i-1,j} [(2-a)\delta_{i,k} \tilde\lambda_j+a\delta_{j,k}\tilde\lambda_i],
\eea
which prove to be consistent with Eqs. (\ref{EF}), provided $a=0$ or $a=2$. Thus, like in the preceding case, one reveals four $\mathcal{N}=2$ models generalizing the non--periodic relativistic Toda lattice.

\vspace{0.5cm}

\noindent
{\bf 4. Geodesic interpretation}\\

\noindent
As was demonstrated in \cite{ABHL}, the rational variant of the Ruijsenaars-Schneider model can be identified with the geodesic equations in a flat space parametrized by special curvilinear coordinates. The non--existence of a {\it metric connection} associated with the hyperbolic Ruijsenaars-Schneider systems was proven in \cite{G}. In this section, we carry out a similar analysis for the relativistic Toda lattice.

We start with the non--periodic case. Rewriting Eqs. (\ref{EOM}) in the form
\be
\ddot x_i+\sum_{j,k=1}^N \Gamma^i_{jk}{\dot x}_j {\dot x}_k=0,
\ee
one obtains the connection coefficients
\bea
\Gamma^i_{jk}=-\frac 12(\delta_{i,j}\delta_{k,i+1}+\delta_{i,k}\delta_{j,i+1})W(x_{i+1}-x_i)+\frac 12(\delta_{i,j}\delta_{k,i-1}+\delta_{i,k}\delta_{j,i-1})W(x_i-x_{i-1}).
\eea
Let us assume that they are derivable from a non--degenerate metric $g_{ij}$
\be
\Gamma^i_{jk}=\frac 12 g^{ip} \left(\partial_j g_{pk}+\partial_k g_{pj}-\partial_p g_{jk} \right),
\ee
where $g^{ij}$ designate the inverse metric components.
Contracting the last formula with $g_{si}$, permuting the indices $(j,s,k)\to(s,k,j)$, and taking the sum, one gets a coupled set of the partial differential equations to fix the metric
\bea\label{sys}
&&
\partial_j g_{sk}=\sum_{i=1}^N (g_{si}\Gamma^i_{jk}+g_{ki}\Gamma^i_{js})=
\nonumber\\[2pt]
&&
\qquad \quad
-\frac 12 g_{sj}[\delta_{k,j+1}W(x_{j+1}-x_j)-\delta_{k,j-1}W(x_j-x_{j-1})]
\nonumber\\[2pt]
&&
\qquad \quad -\frac 12 g_{kj}[\delta_{s,j+1}W(x_{j+1}-x_j)-\delta_{s,j-1}W(x_j-x_{j-1})]
\nonumber\\[2pt]
&&
\qquad \quad -\frac 12 g_{sk}[\delta_{j,k+1}W(x_{k+1}-x_k)-\delta_{j,k-1}W(x_k-x_{k-1})]
\nonumber\\[2pt]
&&
\qquad \quad -\frac 12 g_{sk}[\delta_{j,s+1}W(x_{s+1}-x_s)-\delta_{j,s-1}W(x_s-x_{s-1})].
\eea

Consider three equations belonging to the set (\ref{sys})
\be\label{arg}
\partial_1 g_{11}=0, \quad \partial_2 g_{11}=-W(x_2-x_1)(g_{11}-g_{12}), \quad \partial_1 g_{12}=-\frac 12 W(x_2-x_1) (g_{11}-g_{12}).
\ee
Computing the derivative of the second equation with respect to $x_1$ and taking into account the other two, one gets
\be
\left(\partial_1 W(x_2-x_1)+\frac 12 W(x_2-x_1)^2\right)(g_{11}-g_{12})=0 \quad \Rightarrow \quad g_{11}=g_{12}.
\ee
This is because the first factor entering the leftmost equation is nonzero for the non--periodic relativistic Toda lattice. By repeatedly using the same argument for other components of the metric tensor, one can demonstrate that they all are equal to one and the same constant, $g_{ij}=\mbox{const}$, thus yielding a {\it degenerate} metric. This contradicts to the earlier assumption that $g_{ij}$ is invertible.

The periodic relativistic Toda lattice is treated likewise. Taking into account the boundary conditions (\ref{BC}), the equations of motion (\ref{TL}) can be put into the geodesic form in which the connection coefficients read
\bea
&&
\Gamma^1_{jk}=-\frac 12(\delta_{j,1}\delta_{k,2}+\delta_{k,1}\delta_{j,2})W(x_2-x_1)+\frac 12(\delta_{j,1}\delta_{k,N}+\delta_{k,1}\delta_{j,N})W(x_1-x_N),
\\[2pt]
&&
\Gamma^i_{jk}=-\frac 12(\delta_{i,j}\delta_{k,i+1}+\delta_{i,k}\delta_{j,i+1})W(x_{i+1}-x_i)+\frac 12(\delta_{i,j}\delta_{k,i-1}+\delta_{i,k}\delta_{j,i-1})W(x_i-x_{i-1}),
\nonumber\\[2pt]
&&
\Gamma^N_{jk}=-\frac 12(\delta_{j,1}\delta_{k,N}+\delta_{k,1}\delta_{j,N})W(x_1-x_N)+\frac 12(\delta_{j,N}\delta_{k,N-1}+\delta_{k,N}\delta_{j,N-1})W(x_N-x_{N-1}),
\nonumber
\eea
with $i=2,\dots,N-1$, and $j,k=1,\dots,N$. Considering triples of equations similar to (\ref{arg}), one can again verify that all components of the metric are equal to one and the same constant, which contradicts to the assumption that the metric is invertible.

We thus conclude that, similarly to the hyperbolic Ruijsenaars-Schneider systems, the relativistic Toda models are linked to non--metric connections.

\vspace{0.5cm}

\noindent
{\bf 5. Conclusion}\\

\noindent
To summarize, in this work we have constructed various $\mathcal{N}=2$ supersymmetric generalizations of the relativistic Toda lattice both for the periodic and non--periodic versions. In contrast to the hyperbolic Ruijsenaars-Schneider models, for which only the three--body case has been worked out in full detail \cite{G}, the description above is valid for an arbitrary number of particles. Both the supercharges and the Hamiltonian were shown to have the conventional polynomial form in the fermionic degrees of freedom. A possible geodesic interpretation has been discussed. It was demonstrated that, although the equations of motion of the relativistic Toda lattice can be formally rewritten in the geodesic form, the resulting connection fails to be a metric connection.

Turning to possible further developments, it would be interesting to extend the present study to the $\mathcal{N}=4$ case and to reveal what would be the analog of the Witten--Dijkgraaf--Verlinde--Verlinde equation.
The construction of an off--shell superfield Lagrangian formulation is an interesting open problem.
A generic description of supersymmetric mechanics on spaces endowed with a non--metric connection is a challenge.

One more important question to study concerns quantization of the $\mathcal{N}=2$ models. For the relativistic Toda lattice the conventional strategy is to analyse 
the spectral problem associated with the full set of commuting quantum integrals of motion and to attain the separation of variables (see \cite{R1} for the original consideration and \cite{HM} for a recent alternative treatment). Proceeding to $\mathcal{N}=2$ case, one first has to convert the brackets (\ref{FB}) into those specifying the fermionic creation/annihilation operators and then properly modify the original bosonic quantum integrals of motion in such a way that they commute with the $\mathcal{N}=2$ Hamiltonian in (\ref{SUSY}). This point seems mostly technical. A more severe problem is to prove that the separation of variables is still feasible. We leave these issues for further study.

\vspace{0.5cm}

\noindent{\bf Acknowledgements}\\

\noindent
This work was supported by the Russian Science Foundation, grant No 19-11-00005.

\end{document}